\documentstyle[sprocl,graphicx]{article}
\begin{document}
\noindent

\title{Short-range spin- and pair-correlations: a variational wave-function}

\author{D. van der Marel}

\address{Materials Science Center, University of Groningen \\
Nijenborgh 4, 9747 AG Groningen, The Netherlands\\ and \\
D\'epartement de Physique de la Mati\`ere Condens\'ee \\
Universit\'e de Gen\`eve, Quai Ernest-Ansermet 24, CH1211 Gen\`eve
4, Switzerland }
\date{july 25, 2003}

\maketitle\abstracts{A many-body wavefuction is postulated, which
is sufficiently general to describe superconducting
pair-correlations, and/or spin-correlations, which can occur
either as long-range order or as finite-range correlations. The
proposed wave-function appears to summarize some of the more
relevant aspects of the rich phase-diagram of the high-$T_c$
cuprates. Some of the states represented by this wavefunction are
reviewed: For superconductivity in the background of robust
anti-ferromagnetism, the Cooper-pairs are shown to be a
superposition of spinquantum numbers S=0 and S=1. If the
anti-ferromagnetism is weak, a continuous super-symmetric rotation
is identified connecting s-wave superconductivity to
anti-ferromagnetism.}
\section{Introduction}\label{intro}
The large variety of phenomena due to correlations between
electrons in solids is one of the major challenges of contemporary
physics. Over the past 80 years this subject has formed the arena
of many new developments in theoretical physics. The most common
approach to the many-body problem is, to define in the first step
a model defined as a Hamiltonian, Lagrangian, or action,
describing the most important interactions of the system one tries
to understand. In the second step one tries to solve the
corresponding equations of motion. Apart from a few special cases,
no exact solution is known of the Schr\"odinger equation of a
large number of interacting particles. To circumvent this problem
two different approaches are most frequently used, often in
combination with each other: One is to device an analytical
approximation scheme, which ultimately may provide an approximate
solution. The other is to use computational techniques for a
finite size cluster.

An alternative approach, which has sometimes been quite
successful, is to start with designing a variational many-body
ground-state wave-function with the desired characteristics of
(for example) the pair-correlation function, spin-polarization,
{\em etcetera}. One can than subsequently try to find a model
Hamiltonian which will produce (i) the aforementioned many-body
wave-function as it's ground-state, and (ii) the low energy
excitations which are important for determining the thermal
properties and the various spectral functions all of which can be
measured experimentally.

The variational wave-function approach has been used with great
success for superconductivity\cite{bcs}, the fractional quantum
Hall effect\cite{fqh}, and the resonating valence
bond\cite{rvb1,rvb2,rvb3,rvb4,rvb5}. Other striking examples of
this 'bottom-up' approach are the Affleck-Marsten
flux-phase\cite{fluxphase1,fluxphase2,fluxphase3,fluxphase4,fluxphase5,fluxphase6},
or d-density wave\cite{ddw}, the mixed
anti-ferromagnetic/superconducting state\cite{chen1990}, and the
Gossamer superconducting wavefunction\cite{gossamer1,gossamer2}.

Here we use the many-body wave-function approach to explore the
possibility of (anti)-ferromagnetism, superconductivity, and local
pairs within the same phase-diagram. Our ansatz for the many-body
wavefunction is
\begin{equation}\label{ansatz}
 |\Psi\rangle = C
 \exp{\{\sum_G\sum_k f(k,G)c^{\dagger}_{k\uparrow}
 c^{\dagger}_{-k+G\downarrow}\}} |0\rangle
\end{equation}
where $C$ is a normalization constant. The function $f(k,G)$ is a
distribution function of the center of mass momentum ($G$) and the
relative momentum ($2k$) of pairs of electrons. Below we will see,
that $f(k,G)$ represents several types of different correlations
within the interacting electronic system. Moreover Eq.
\ref{ansatz} is sufficiently general to capture, besides less
familiar states of matter, several well known states of matter,
such as superconductivity, anti-ferromagnetism, resonating valence
bond states, and the non-interacting electron gas. It therefor may
be a useful starting point for the rich phase-diagram of the
high-T$_c$ cuprates, and of the heavy-fermion superconductors. In
the subsequent discussion we will frequently refer to the
projection of Eq.\ref{ansatz} on the subspace with $2N$ electrons
\begin{equation}\label{variational}
 |\Psi_{2N}\rangle = C
 \{ \sum_G\sum_k f(k,G)c^{\dagger}_{k\uparrow}
 c^{\dagger}_{-k+G\downarrow} \}^N |0\rangle
\end{equation}
This expression is a generalization of the many-body wavefunction
describing the simultaneous occurrence of superconductivity and
anti-ferromagnetism\cite{chen1990} to arbitrary distributions,
$f(k,G)$, of the center of mass momentum $G$.

\section{Superconductivity}
Let us first consider the special case, that the distribution
function $f(k,G)$ is a Dirac $\delta$-function of the
center-of-mass momentum: $f(k,G)=\delta(G,0)\psi_{k}$.
\begin{equation}\label{variational1}
|\Psi_{2N}\rangle = C \hat{P}_{2N} \exp{\{A_0^{\dagger}\}}
|0\rangle
\end{equation}
where $\hat{P}_{2N}$ is the projection on states with $2N$
electrons, and
\begin{equation}\label{A0}
  A_0^{\dagger}\equiv\sum_k \phi_k c^{\dagger}_{k\uparrow}c^{\dagger}_{-k\downarrow}
\end{equation}
creates a pair of electrons with zero center-of-mass momentum. The
exponential function of $A_0^{\dagger}$ appearing in
Eq.\ref{variational1}, can be expanded in a Taylor's series. The
term containing $2N$ electrons
\begin{equation}\label{variational2}
  |\Psi_{2N}\rangle = \{A_0^{\dagger}\}^N |0\rangle
\end{equation}
is similar to a Bose-Einstein condensate of $N$ composite bosons,
each consisting of a pair of electrons. In a superconductor this
state is realized if we cool an isolated lump of material with
known number of electrons below the phase transition. Bardeen,
Cooper and Schrieffer considered a phase-coherent superposition of
states containing different numbers of pairs:
\begin{equation}\label{BCS1}
  |\Psi_{\phi}\rangle = C \sum_N e^{i\phi N}|\Psi_{2N}\rangle
\end{equation}
which can be shown to be equivalent to the state
\begin{equation}\label{BCS2}
  |\Psi_{\phi}\rangle = \prod_k\left(u_k+e^{i\phi}v_kc^{\dagger}_{k\uparrow}c^{\dagger}_{-k\downarrow} \right) |0\rangle
\end{equation}
where $\psi_k=v_k/u_k$, and $|v_k|^2+|u_k|^2=1$. In the limit
case, where $|v_k|^2\rightarrow 1$ for $|k| \le k_F$ and
$|v_k|^2\rightarrow 0$ for $|k| > k_F$, this product is just the
ground-state of the non-interacting electron gas
\begin{equation}\label{fermi-gas}
  |\Psi_{0}\rangle = \prod_{k,\sigma} \theta_k^F c^{\dagger}_{k\sigma} |0\rangle
\end{equation}
where the occupation number $\theta_k^F$ is 0 or 1 depending on
whether the state is above or below the Fermi level respectively.
\section{Anti-ferromagnetism}
Another well-documented limit of Eq. \ref{variational} is
presented by the anti-ferromagnetic state. To be specific, we
consider a square lattice in two space-dimensions, with
site-alternating up- and down polarization. The anti-ferromagnetic
Bragg vector is {\bf Q}=$(\pi,\pi)$ in this case, and the
operators which create an electron on the two sublattices are
\begin{equation}\label{opab}
  \begin{array}{l}
    a^{\dagger}_{k\sigma} = \frac{1}{\sqrt{2}}
    \left(c^{\dagger}_{k\sigma}-c^{\dagger}_{k+Q\sigma}\right)\\
    b^{\dagger}_{k\sigma} = \frac{1}{\sqrt{2}}
    \left(c^{\dagger}_{k\sigma}+c^{\dagger}_{k+Q\sigma}\right)\\
  \end{array}
\end{equation}
respectively.
The single particle operators of the anti-ferromagnetic state are
\begin{equation}\label{AFeigen}
  \begin{array}{lllllr}
  \epsilon_{k}^{-}: &
  \alpha^{\dagger}_{k\uparrow}=&\mu_k c^{\dagger}_{k\uparrow}-\nu_kc^{\dagger}_{k+Q\uparrow}& \mbox{and}
  &\beta^{\dagger}_{k\downarrow}=&\nu_k c^{\dagger}_{k\downarrow}+\mu_kc^{\dagger}_{k+Q\downarrow}\\
  \epsilon_{k}^{+}: &
  \beta^{\dagger}_{k\uparrow}=&\nu_k c^{\dagger}_{k\uparrow}+\mu_kc^{\dagger}_{k+Q\uparrow} &\mbox{and}
  &\alpha^{\dagger}_{k\downarrow}=&\mu_k c^{\dagger}_{k\downarrow}-\nu_kc^{\dagger}_{k+Q\downarrow}\\
  \end{array}
\end{equation}
where the coefficients satisfy $\mu_k =\mu_{-k}$, $\nu_k =
\nu_{-k}$, $\mu_{k+Q} =\nu_{k}$, $\nu_{k+Q} =\mu_{k}$, and
$|\mu_k|^2+|\nu_k|^2=1$. The energies $\epsilon_{k}^{\pm}$ are
determined by the details of the energy-momentum dispersion of the
$c^{\dagger}_{k\sigma}$-operators and by the interactions giving
rise to the anti-ferromagnetism. For simplicity we consider here
the case where only one of the two sub-bands is occupied, which
assumes that the anti-ferromagnetic splitting is very large. The
ground-state wave-function is obtained by partially or fully
occupying the lowest band
\begin{equation}\label{AF1}
  |\Psi_{AF}\rangle = \prod_k\theta_k^F\alpha^{\dagger}_{k\uparrow} \beta^{\dagger}_{k\downarrow} |0\rangle
\end{equation}
The states with momentum $k$ and $-k$ are degenerate, so we may
replace $\alpha^{\dagger}_{k\uparrow}
\beta^{\dagger}_{k\downarrow}$ with $\alpha^{\dagger}_{k\uparrow}
\beta^{\dagger}_{-k\downarrow}$ in the above product. Because each
electron-state can only be created once, this is equivalent to
\begin{equation}\label{AF2}
  |\Psi_{AF}\rangle =  \{2^{-1/2}\sum_k\theta_k^F\alpha^{\dagger}_{k\uparrow} \beta^{\dagger}_{-k\downarrow}\}^N |0\rangle
\end{equation}
The derivation of Eq. \ref{AF2} employs the property that
$\alpha^{\dagger}_{k\uparrow}$ and $\beta^{\dagger}_{k\downarrow}$
are orthogonal to each other, consequently all two-particle
operators under the summation commute with each other. In the
first (reduced) Brillouin-zone of the anti-ferromagnetic state the
occupation function $\theta_k^F$ is either 1 or 0. In Eq.
\ref{AF2} and in later similar expressions the summation over $k$
refers to the paramagnetic (extended) Brillouin-zone. The factor
$2^{-1/2}$ compensates for the double counting. However,
$\alpha^{\dagger}_{k\sigma}=-\alpha^{\dagger}_{k+Q\sigma}$ and
$\beta^{\dagger}_{k\sigma}=\beta^{\dagger}_{k+Q\sigma}$. To avoid
the complete cancellation of terms originating from the first and
second reduced Brillouin-zone, it is therefor important to define
$\theta_{k+Q}^F=-\theta_{k}^F$ for $k$ in the second reduced
Brillouin-zone.

To see the relation with Eq. \ref{variational} we substitute for
$\alpha^{\dagger}_{k\uparrow}$ and
$\beta^{\dagger}_{-k\downarrow}$ the original
$c^{\dagger}_{k\sigma}$-operators in the paramagnetic
Brillouin-zone  using Eq. \ref{AFeigen}. We then decompose the
summation over $k$ in separate terms corresponding to different
quantum numbers for the spin and the center-of-mass momentum
\begin{equation}\label{AF3}
\begin{array}{l}
  \sum_k\theta_k^F\alpha^{\dagger}_{k\uparrow} \beta^{\dagger}_{-k\downarrow}
   =   2\sum_k \theta_k^F \left(\mu_k \nu_k c^{\dagger}_{k\uparrow}c^{\dagger}_{-k\downarrow}
   +  \mu_{k}^2
   c^{\dagger}_{k\uparrow}c^{\dagger}_{-k+Q\downarrow}\right)
   \\
   =\sum_k f^e_0(k) c^{\dagger}_{k\uparrow}c^{\dagger}_{-k\downarrow}
   +  \sum_{k'} (g^e(k')+g^o(k)) c^{\dagger}_{k'+Q/2\uparrow}c^{\dagger}_{-k'+Q/2\downarrow}
\end{array}
\end{equation}
where
\begin{equation}\label{AF3b}
\begin{array}{rll}
  &f^e_0(k)=&\theta_k^F\left(2\mu_k \nu_k\right) \\
  &g^e(k')=&\theta_{k'+Q/2}^F\left(\mu_{k'+Q/2}^2-\nu_{k'+Q/2}^2 \right) \\
  &g^o(k')=&\theta_{k'+Q/2}^F
\end{array}
\end{equation}

This concludes the proof, that Eq. \ref{variational} is
sufficiently general to also contain the anti-ferromagnetic state
as one of the possibilities. The distribution-function of the
center-of-mass momentum, $f(k,G)$, is a superposition of two
$\delta$-functions: The first at $G=0$ is an even function of $k$
and therefor creates a pair with spin quantum number zero. The
second $\delta$-function at the anti-ferromagnetic Bragg-vector
$G=(\pi,\pi)$ contains two contributions with different spin
quantum-numbers. Since
$\theta_{-k'+Q/2}=\theta_{k'-Q/2}=-\theta_{k'+Q/2}$, and
$\mu_{-k'+Q/2}^2=\mu_{k'-Q/2}^2=\nu_{k'+Q/2}^2$, the functions
$g^e(k')$ and $g^o(k')$ are even and odd functions of $k'$
respectively. Hence the first of the two terms with center-of-mass
momentum $Q$ corresponds to a S=0 state. The other term, because
it is an odd function of $k'$, corresponds to the $m_S=0$ member
of the ($S=1$) spin-triplet.

A well-known property of the anti-ferromagnetic state is the
broken SU(2) symmetry. This symmetry breaking is illustrated by
the fact, that Eq. \ref{AF3} corresponds to a superposition of
pair-states with different spin quantum numbers ({\em i.e.} S=0
and S=1). Eq. \ref{AF3} is also a superposition of pair-states
with different center-of-mass quantum numbers, and this
illustrates the fact that the anti-ferromagnetic state also breaks
the discrete translational invariance of the lattice.

The spin-polarization depends on the value of $\mu_k^2=1-\nu_k^2$.
For example $\mu_k=\nu_k=1/\sqrt{2}$ corresponds to full
spin-polarization, where the electrons move either completely in
sublattice $a$ with spin $\uparrow$, or in sublattice $b$ with
spin $\downarrow$. From Eq. \ref{AF3b} we see that the fully
polarized anti-ferromagnetic state can be regarded as a
simultaneous condensation in a singlet state with momentum $G=0$
and a triplet with momentum $G=(\pi,\pi)$, both having the same
amplitude.

\section{Superconductivity and anti-ferromagnetism}
The results of this section and the previous one can be summarized
by a diagram displaying the center-of-mass distribution function
$f(k,G)$ averaged over the relative coordinate $k$,
$F(G)\equiv\sum_k|f(k,G)|^2$. In Fig. 1 this function is displayed
for the superconducting state and the anti-ferromagnetic state.
\begin{figure}\label{lro}
  \centerline{\includegraphics[width=7cm,clip=true]{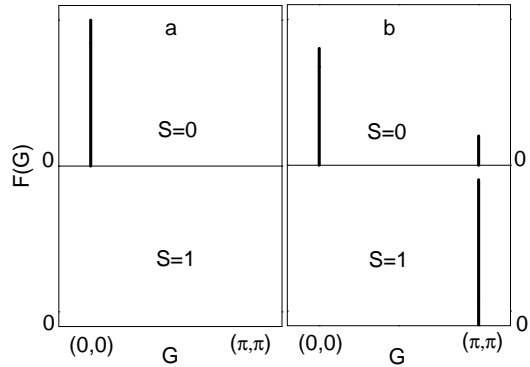}}
  \caption{ Distribution of the center-of-mass momentum for
  {\it a)} a BCS-superconductor
  {\it b)} an anti-ferromagnet. The positive (negative) values correspond to
  the singlet (triplet) channels. The vertical bars indicate
  $\delta$-functions, the area of which is represented by the length of the
  bar.}
  \end{figure}
We point out two important aspects of the BCS-wavefunction of the
superconducting state:

1) The internal structure of the Cooper-pairs ({\em i.e.} the
pairing symmetry and the degree of localization of the pairs) is
completely determined by $v_k$ and $u_k$ in Eqs. \ref{BCS1} and
\ref{BCS2}: The function $\psi_k=v_k/u_k$ is the Fourier-transform
of the wavefunction describing the relative coordinate of the
electrons forming a Cooper-pair\cite{nozieres-schmidt-rink}.

2) All pairs are condensed in a state with zero center-of-mass
momentum $G=0$, regardless of the details of the internal
structure of the pairs.

Although the long-range anti-ferromagnetic state is characterized
by two $\delta$-functions for the center-of-mass momentum $G$,
Eq.\ref{AF3} does not correspond to a true superconducting state,
because the occupation function $\theta_k^F$ has a sharp cut-off
at the Fermi-momentum.

\subsection{Superconductivity in the background of
anti-ferromagnetism} We first consider the situation, where the
anti-ferromagnetism has been stabilized on a high energy scale,
and superconductivity is a relatively weak phenomenon. Under those
conditions it is a good approximation to assume that the onset of
superconductivity does not affect the anti-ferromagnetic order
parameter. The simultaneous occurrence of superconductivity and
anti-ferromagnetism requires that $\theta_k^F$ is replaced with
the smooth function $v_k/u_k$, as usual in a BCS-type
superconductor. In principle this smooth function may be different
for the three different terms with differing quantum numbers, and
the corresponding many-body wavefunction was described by Chen
{\em et al.}\cite{chen1990}, but in the fully polarized
anti-ferromagnet we expect a single distribution function, so that
the superconducting/anti-ferromagnetic wavefunction can be
transformed to up- and down-spin sublattice-operators using Eq.
\ref{opab}
\begin{equation}\label{AF1b}
  |\Psi_{AF}\rangle = \prod_k\frac{v_k}{u_k}a^{\dagger}_{k\uparrow} b^{\dagger}_{k\downarrow} |0\rangle
\end{equation}
Interestingly in this example the Cooper-pairs are a superposition
of an S=0 and an S=1 spin-state with equal amplitudes for the S=0
and S=1 contributions. In other words, the superconducting state
of a fully polarized anti-ferromagnet is a condensate with a
broken SU(2) symmetry, with Cooper-pairs which have one spin-up
electron on sublattice a, and one spin-down electron on sublattice
b. In the partially polarized anti-ferromagnet the amplitude of
the S=1 contribution is different from the S=0 weight a
$Q=(\pi,\pi)$. The degree of polarization of the Cooper-pairs
depends on the details of the pairing-mechanism and the degree of
polarization of the underlying anti-ferromagnetic state. For
vanishing spin-polarization SU(2)-symmetry should be restored, and
the Cooper-pairs are either in an S=1 state or in an S=0 state,
depending on the parity of $v_k/u_k$.

\section{Supersymmetric rotations and SO(5)}
A different situation arises, when the anti-ferromagnetic
correlations and the superconducting correlations are stabilized
on comparable energy scales. In this case the anti-ferromagnetism
can be partially or totally suppressed when superconducting order
sets in and vice-versa. An elegant approach to this phenomenon is
the group-theory, which uses the super-symmetric SO(5) extension
of the direct product of U(1) corresponding to the superconducting
phase, and of SO(3) corresponding to the anti-ferromagnetic order
parameter\cite{so5}. In this so-called SO(5) theory, the
anti-ferromagnetic state and the d-wave superconducting state are
regarded as two different projections of a generalized higher
dimensional order parameter. It is tempting to regard Eq.
\ref{AF3} as a manifestation of precisely this SO(5) symmetry,
namely, that the d-wave superconducting condensate can be
transformed into anti-ferromagnetic state by adiabatically adding
the additional triplet condensate at $G=(\pi,\pi)$. However, in
Eq. \ref{AF3} the singlet at $G=0$ has s-wave symmetry if the
anti-ferromagnetic state is of the conventional variety. Hence,
although our wave-function {\em does} support a super-symmetric
rotation from anti-ferromagnetism to s-wave pairing by mixing in a
triplet-condensate at $G=(\pi,\pi)$, a similar construction
connecting d-wave pairing to conventional anti-ferromagnetism
appears to be absent\cite{gas}. In principle we could postulate an
unconventional anti-ferromagentic order parameter $\mu_k$ which
has a d-wave symmetry with nodes along the diagonal directions.
Such an order parameter transforms to a d-wave superconducting
order parameter if we reduce the singlet and triplet condensates
at $G=(\pi,\pi)$ to zero.
\section{Local pairs}
It has often been speculated that a remnant of the Cooper-pairs
may exist even if the material is not superconducting. This would
require that those pairs are not condensed in the same state. The
variational wave-function proposed by Bardeen, Cooper and
Schrieffer does not include this possibility. On the other hand
the situation sketched in Fig. 1a can be easily generalized to
describe a paired state with no long range superconducting order,
by replacing the Dirac $\delta$-function of the center-of-mass
coordinate with the Lorentzian $~1/(1+l^2|G|^2)$. The width of
this distribution, $1/l$ then corresponds to the inverse of length
over which the pairs maintain the phase-coherence. A transition to
the superconducting state as a function of temperature, or another
control parameter, is in this scenario characterized by the
appearance of the Dirac $\delta$-function at the origin. Note,
that for the existance of superconductivity it is not necessary
that the entire distribution collapses into the $\delta$-function,
not even at $T=0$. This scenario, depicted in Fig. 2, is
reminiscent of the experiments with Bose-Einstein condensation.
Yet, the pairs in this wave-function are not real bosons. In
principle it is not even required that they are strongly localized
in space, although the conditions giving rise to this type of
phase transition would typically cause the pairs to be rather
small.
\begin{figure}\label{srso}
  \centerline{\includegraphics[width=7cm,clip=true]{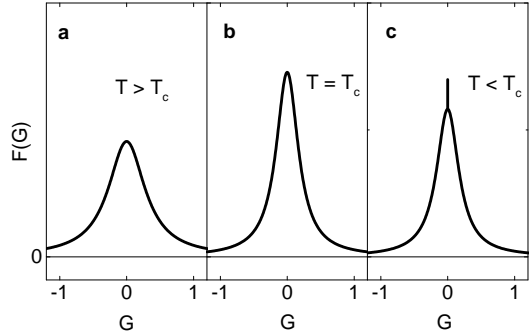}}
  \caption{Non-BCS scenario for the evolution from the normal to the
  superconducting state. Panels {\it a} and {\it b} correspond to a situation
  where pairs are already formed, but no condensation has yet
  occurred. The superconducting state is only reached
  when the $\delta$-function gains a finite intensity, as
  indicated in panel {\it c}.}
\end{figure}
States of the type displayed in Fig. 2a have a simple
representation in real space. Starting from Eq.\ref{variational},
we insert the real-space representation of the creation operators
$c^{\dagger}_{k\sigma}=\sum_{\bf R} e^{ikR}
c^{\dagger}_{j\sigma}$, and obtain
\begin{equation}\label{rvb1}
 |\Psi_{2N}\rangle = C
 \{ \sum_{m,n} \tilde{f}(r_m,r_n) c^{\dagger}_{m\uparrow}
 c^{\dagger}_{n\downarrow} \}^N |0\rangle
\end{equation}
where $\tilde{f}(r_m,r_n)\equiv \sum_{k,G}e^{i(k\cdot
(r_m-r_n)+G\cdot r_n)}f(k,G)$ is the double Fourier-transform of
the momentum distribution function. A state with near-neighbor
resonating bonds is constructed by defining
\begin{equation}\label{rvb2a}
 \tilde{f}(r_m,r_n)=
 e^{-|r_m+r_n-R_0|/(2l)} \delta_{<m,n>} \left\{(x_{m}-x_{n})+i(y_m-y_n)\right\}^L
\end{equation}
where $l$ measures the range of the RVB correlations measured from
an arbitrary reference point ($R_0$) where the correlations are
maximal, $\delta_{<m,n>}$ selects out nearest-neighbour bonds, and
the last factor selects the state with relative angular momentum
L. A Gutzwiller projection is still required if we want to exclude
the doubly occupied sites. The representation in k-space is
\begin{equation}\label{rvb2b}
 |\Psi_{2N}\rangle \approx C \hat{P}
 \left\{ \sum_{G,k} e^{i\phi(G)}\frac{\cos(k_x)\pm\cos(k_y)}{1+l^2|G|^2}c^{\dagger}_{k+G/2\uparrow}
 c^{\dagger}_{-k+G/2\downarrow} \right\}^N |0\rangle
\end{equation}
where the $+/-$ sign refers to $L=0$ or $L=2$ angular momentum
states respectively, $\phi(G)$ is a random phase, and $\hat{P}$ is
the Gutzwiller projection operator. The true RVB
state\cite{rvb1,rvb2,rvb3,rvb4,rvb5} corresponds to taking the
limit $l\rightarrow\infty$ in the above expression. In this limit
this state becomes equivalent to a BCS-type superconducting
wave-function with either extended s-wave symmetry or d-wave
symmetry, with the two paired holes on a nearest-neigbour
distance. In a superconductor with a small gap, the $k$-dependent
factor inside the curly brackets has a strong k-dependence,
corresponding to singlet-bonds on a range much longer than a
nearest neighbor distance in $\tilde{f}(r_m,r_n)$.
\section{Finite range magnetic correlations}
Likewise a wavefunction corresponding to a state of short-range
magnetic correlations can be easily constructed by replacing the
Dirac $\delta$-functions in Fig. 2b with distribution functions
with a finite width. The corresponding structure sketched in Fig.
3a corresponds to the magnetic structure-factor measured with
neutron scattering, and the width of the distribution is inversely
proportional to the anti-ferromagnetic correlation length. Like
for the superconducting state, long-range anti-ferromagnetic order
is characterized by the condensation of part of the spectral
weight into the $\delta$-function at $G=(\pi,\pi)$.
\begin{figure}\label{srafo}
  \centerline{\includegraphics[width=7cm,clip=true]{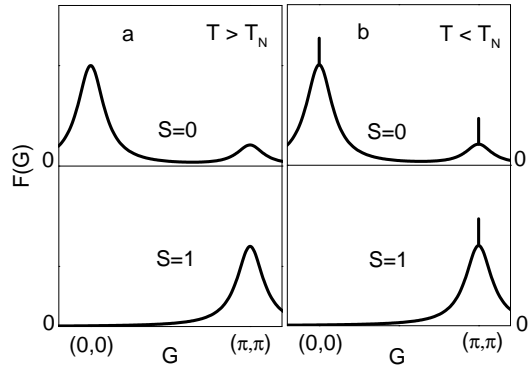}}
  \caption{Scenario for the evolution from a state with short range anti-ferromagnetic
  order (panel {\it a})to the N\'eel state with long range order (panel {\it b}).}
\end{figure}
By shifting the peak in the distribution function from $(\pi,\pi)$
to another value of the wavevector $G$, the formalism can be used
to describe short- or long-range incommensurate spin-density
waves.

\section{Summary and conclusions}
We have generalized earlier expressions of the many-body
wavefuction\cite{bcs,rvb2,chen1990} to a formula, Eq.\ref{ansatz},
which envelopes many known or suspected types of spin- and pairing
order of the high T$_c$ cuprates and of some of the heavy fermion
superconductors. We have used this approach to show that
anti-ferromagnetism can be regarded as the simultaneous
condensation of S=0 singlets at $G=0$ and $G=Q$, and of S=1
triplets at $G=Q$ where $Q$ is the anti-ferromagnetic
Bragg-vector. The possibility of superconductivity and
anti-ferromagnetism occurring simultaneously can be easily
described this way. In principle the calculation of
matrix-elements of the Hamiltonian, as well as most operators
corresponding to experimentally measurable quantities, is
unproblematic at least in the examples where the center-of-mass
momentum, $G$, has a single discrete value. The general case
presented by the proposed wave-function allows $G$ to be
distributed over a finite range of $k$-space. In the latter case
it appears to be more problematic to obtain analytical expressions
of the aforementioned matrix-elements. The latter forms, as yet,
the main obstacle for using Eq. \ref{ansatz} in a variational
calculation of the ground state and the excitation spectrum. If
this obstacle can be taken, this could offer a straightforward
analytical approach to the rich phase diagram of the high T$_c$
cuprates.

The work presented in this paper was developed during a long term
collaboration with M. J. Rice. His pedagogical teachings have
largely influenced the present paper, written in his memory. It
has been a great privilege to collaborate with Michael.


\end{document}